\documentclass[letterpaper,aps,prl,amsmath,amssymb,floatfix,twocolumn,showpacs,superscriptaddress]{revtex4-1}
\usepackage{graphicx}
\usepackage{hyperref}
\usepackage{dcolumn}
\usepackage{bm}
\usepackage{xcolor}
\hypersetup{
    colorlinks,
    linkcolor={red!80!black},
    citecolor={blue!50!black},
    urlcolor={blue!80!black}
}

\usepackage{epstopdf}
\usepackage{times}
\usepackage{mathptmx}

\begin{document}

\title{Efficient detection of contagious outbreaks in massive metropolitan encounter networks}
\author{Lijun Sun}
\email[Corresponding author:  ]{sunlijun@nus.edu.sg}
\affiliation{Future Cities Laboratory, Singapore-ETH Centre for Global Environmental Sustainability (SEC), 138602, Singapore}
\affiliation{Department of Civil \& Environmental Engineering, National University of Singapore, 117576, Singapore}
\affiliation{Media Laboratory, Massachusetts Institute of Technology, Cambridge, Massachusetts 02139, USA}
 \author{Kay W. Axhausen}
\email{axhausen@ivt.baug.ethz.ch}
\affiliation{Future Cities Laboratory, Singapore-ETH Centre for Global Environmental Sustainability (SEC), 138602, Singapore}
\affiliation{Institute for Transport Planning and Systems (IVT), Swiss Federal Institute of Technology, Z\"{u}rich, 8093, Switzerland}

\author{Der-Horng Lee}
\email{dhl@nus.edu.sg}
\affiliation{Department of Civil \& Environmental Engineering, National University of Singapore, 117576, Singapore}

\author{Manuel Cebrian}
\email{manuel.cebrian@nicta.com.au}
\affiliation{National Information and Communications Technology Australia, University of Melbourne, Victoria 3010, Australia}

\begin{abstract}
Physical contact remains difficult to trace in large metropolitan networks, though it is a key vehicle for the transmission of contagious outbreaks. Co-presence encounters during daily transit use provide us with a city-scale time-resolved physical contact network, consisting of 1 billion contacts among 3 million transit users. Here, we study the advantage that knowledge of such co-presence structures may provide for early detection of contagious outbreaks. We first examine the ``friend sensor'' scheme - a simple, but universal strategy requiring only local information - and demonstrate that it provides significant early detection of simulated outbreaks. Taking advantage of the full network structure, we then identify advanced ``global sensor sets'', obtaining substantial early warning times savings over the friends sensor scheme. Individuals with highest number of encounters are the most efficient sensors, with performance comparable to individuals with the highest travel frequency, exploratory behavior and structural centrality. An efficiency balance emerges when testing the dependency on sensor size and evaluating sensor reliability; we find that substantial and reliable lead-time could be attained by monitoring only 0.01\% of the population with the highest degree.
\end{abstract}


\maketitle

\section{Introduction}

Digital traces generated by citizens, during their commute across metropolitan transportation networks are helping answer long-standing questions on topics from individual mobility to collective interaction patterns. A series of landmark papers examining multiple large-scale digital traces has shifted the understanding of individual mobility patterns from random to highly structured and predictable \cite{Brockmann2006,Gonzalez2008,Song2010,deMontjoye2013,Sun2013}. This has important implications in urban dynamics and epidemiology, particularly as the reproducible structure of metropolitan face-to-face encounters does significantly shape the spatial-temporal dynamics of disease spreading \cite{Balcan2009,Salathe2010,Funk2010}. Therefore, advances in deciphering metropolitan encounter patterns play an important role in detection and mitigation of contagious outbreaks \cite{Huerta2002,Eubank2004,Wang2009}.

In detecting and containing contagious outbreaks, it is crucial to identify ``super-spreaders'', as they may provide significant lead indicators for the early response of public health agencies \cite{Galvani2005,Kitsak2010}. To measure an individual's importance in spreading processes, various centrality measures, such as degree, betweenness, closeness \cite{Freeman1979}, $k$-shell index \cite{Kitsak2010} and activity potential \cite{Perra2012} have been applied to theoretical diffusion models. Recent empirical works have confirmed the importance of these diverse measurements in real-world diffusion processes \cite{Kitsak2010,Lv2011,Aral2012,Bauer2012,Borge2012,Perra2012,Smieszek2013}. To obtain such measurements, full knowledge about the contact network structure is usually required; however, other than simulating human interaction at this level of resolution \cite{Eubank2004,Balcan2009,Bajardi2011}, mapping such structure from real-world physical contact processes could be expensive to collect, computationally costly, laborious in the filtering of spurious connectivity, and privacy-sensitive \cite{Christakis2010,Cattuto2010,Salathe2010,Stehle2011}. This has been particularly true for large metropolitan contact networks, where the availability of citywide datasets is still limited \cite{Kuiken2000,Gilbert2007}.

Disease monitoring is extremely costly, privacy sensitive, and involves enormous technical difficulties. A low-cost contact network structure constructed from transit use may provide a way to design efficient monitoring strategies using a small fraction of the population. In this work, we examine the largest metropolitan encounter dataset collected to date - travel smart card data from all of Singapore's bus users, covering approximately 3 million users during 1 week. Using one week's tapping-in/tapping-out data collected from public transit services in Singapore, we built a large-scale high-resolution physical contact network. In a recent study based on this dataset, we demonstrated that physical encounters display a significant degree of temporal regularity and these rhythmic interactions form a large-scale spatial-temporal contact network, spanning all of Singapore for the whole week \cite{Sun2013}. The study emphasizes that encounters at this fine-grained scale are also very structured, and far from random. If the former study identified the global behavioral properties that generate this citywide co-presence network, our present study tries to identify the key individuals' network properties that can be exploited to combat the spread of infectious disease.

As an alternative to constructing a global structure of contact networks, recent research exhibits an increasing interest in applying crowd-sourcing as a potential strategy to detect contagious outbreaks, from using declared ``friends as sensors'', to aggregated search engine queries, to social media \cite{Ginsberg2008,Chan2010,Garcia2012,Shaman2012,Hodas2013,Salathe2013}. Although these methods proposed are based on simple principles and require only small slices of information, they also show great advantages in providing early warning. Still, interesting questions remain in comparing the possible gains of using full knowledge vs. local methods in an epidemiological city-wide scenario. We perform such study in this high-resolution network, as a first evidence of its kind at a population and metropolitan level.

\begin{figure*}[t]
\begin{center}
\centerline{\includegraphics[scale=0.65]{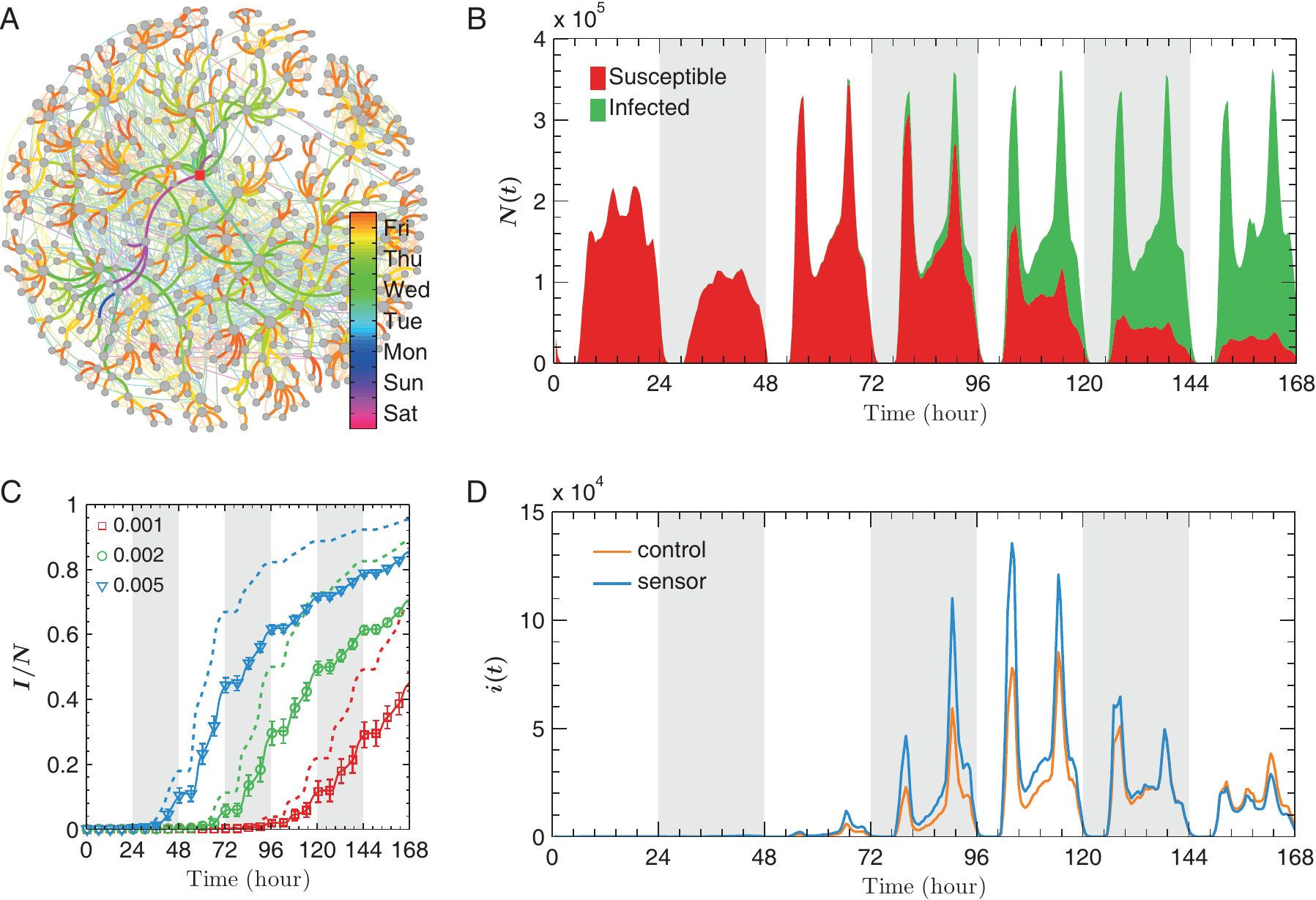}}
\caption{Modeling contagious outbreaks in a city-scale physical contact network. (A) Simulated infection processes from one infectious individual (red square). The encounter network is drawn in two layers: effective infection path (solid links in full color) and the remainder physical encounters (thin links with opacity). (B) Temporal (hourly) change of susceptible and exposed people and infected people across the population. The results come from one simulation with contagion rate $\beta =0.0015$, demonstrating how transit users become infected from day to day. (C) Temporal ratio of infected and susceptible from 20 simulations with different contagion rate $\beta$. The solid curves show average ratios $\left\langle {{I}_{P}}/{{N}_{P}} \right\rangle $ over 20 runs and error bars indicate standard deviation. The dashed curves show the average trend of infected ratio $\left\langle {{I}_{S}}/{{N}_{S}} \right\rangle $ of the 1\% friend sensors. Lead-time can be estimated by checking time difference when $\left\langle {{I}_{P}}/{{N}_{P}} \right\rangle $ reaches certain value. (D) Number of hourly infection incidences during the week, from the same simulation run as in panel (B). The orange dashed curve and the blue solid curve illustrate the temporal variation in population $C$ and the selected friend group $S$, respectively.}
\label{fig1}
\end{center}
\end{figure*}

\section{Results}

To explore the dynamics of city-scale contagious outbreaks, we applied a general Susceptible-Exposed-Infected (SEI) model \cite{Anderson1992} to simulate the spreading processes (see Methods and Supplementary Note 1). Briefly, a simulation run is initialized with ten infectious people (as index cases), who are selected randomly among all transit users on Saturday. In the temporal weighted physical encounter network (with each contact as an edge and its duration as weight), an infectious individual $i$ will transmit disease to neighbor $j$ with probability ${{p}_{ij}}=\beta {{d}_{ij}}$ per 20 seconds (contagion rate $\beta $ is a universal parameter across the population and ${{d}_{ij}}$ represents encounter duration; see Fig.~\ref{fig1}A for example). Once a susceptible individual get exposed, he/she becomes infectious after 2 hours, starting to spread the disease to other susceptible people. As almost all transit journeys are shorter than 2 hours, the introduction of this exposure stage prevents one from getting infected and then infects others directly during the same journey (which will significantly boost the spreading as instantaneous networks for a vehicle is always fully connected). Note that $\beta =0.003$ is used in a high-resolution contact network in Ref \cite{Salathe2010,Smieszek2013}; we apply comparable values in our simulations. The full temporal resolution enables us to simulate the spreading processes during the whole week based on the proposed scheme for detecting contagious outbreaks, by registering infection time and transmission pathway on individual levels (Fig.~\ref{fig1}B).

As mentioned above, a simple, but effective strategy for early detecting contagious outbreaks without mapping the detailed structure of a social network is to find friend sensors from the population \cite{Christakis2010}. The inherent principle behind this method: a randomly selected ``friend'' (neighbor; in a friend group) of one vertex (in a control group) has higher degree on average when the network has a heterogeneous degree distribution, implying that friend group is more central than the control group (or the population as a whole). This is commonly referred as the ``friendship paradox''; your friends have more friends than you do \cite{Feld1991}. However, as social links initiated by physical encounters with strangers display a significant degree of heterogeneity, it remains unclear whether the friend sensor scheme - obtained from a static network structure - works in temporal spreading processes. Hence, to assess performance of the friend sensor scheme, we conducted multiple simulation experiments with different contagion rates $\beta $. In each simulation, we first select 1\% individuals from population $P$ randomly as a control set $C:=\left\{ {{c}_{i}}|{{c}_{i}}\in P \right\}$; the corresponding sensor group $S$ is composed of randomly selected neighbors of each individual in $C$ ($S:=\left\langle {{s}_{i}}|{{s}_{i}}\in N\left( {{c}_{i}} \right),{{c}_{i}}\in C \right\rangle $, and $N\left( {{c}_{i}} \right)$ is a neighbor set of individual ${{c}_{i}}$). Note that $S$ is a list instead of a set since an individual might be selected repeatedly from different $N\left( {{c}_{i}} \right)$. After obtaining results from 20 simulations, we measured the average infected ratio $\left\langle {{I}_{S}}/{{N}_{S}} \right\rangle $ of the sensor groups and $\left\langle {{I}_{P}}/{{N}_{P}} \right\rangle $ of the whole population temporally, finding that friend sensors have large lead-times (Fig.~\ref{fig1}C). Given the heterogeneous individual participation and size of the time window \cite{Krings2012}, spreading exhibits a linear increase - instead of a saturation process - after the explosive stage.

In Fig.~\ref{fig1}D, we show the temporal change of infection incidence $i\left( t \right)$ from the same simulation as Fig.~\ref{fig1}B. The sensor group $S$ is obtained by the same selection scheme; however, in this case, the control group $C$ is the whole population. Together with Fig.~\ref{fig1}C, we found that spreading in $S$ not only happens earlier, but also faster than in the whole population, suggesting that the lead-time also varies with time (or infected ratio; see Supplementary Note 3 and Supplementary Fig.~\ref{figs1} and~\ref{figs2}). Notably, although the temporal structure is not used in finding sensors, the friend sensor scheme is still efficient in early detecting outbreaks in our simulation experiments.

\begin{figure}[tbp]
\begin{center}
\centerline{\includegraphics[scale=0.65]{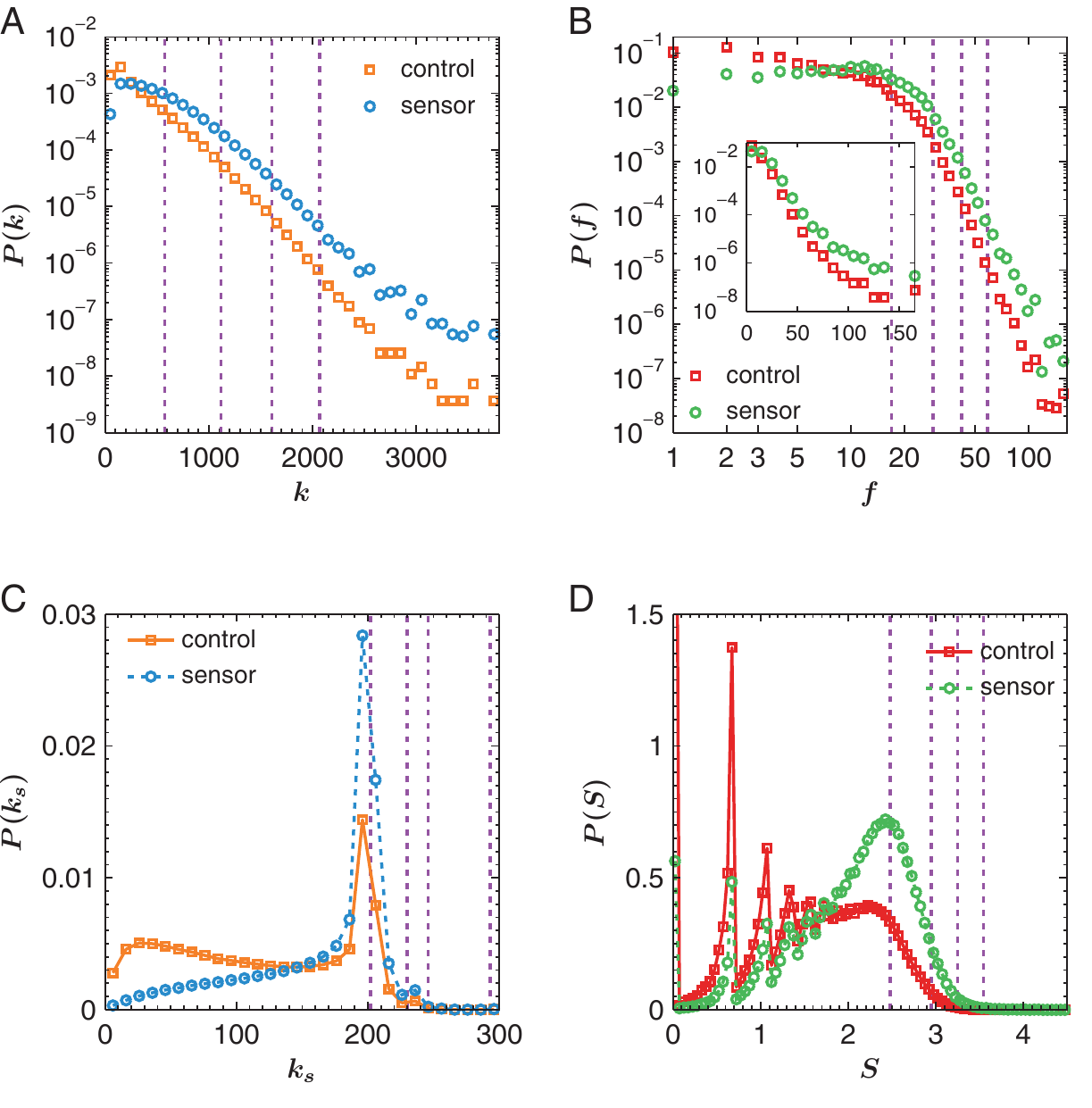}}
\caption{The ``friendship paradox'' exhibited in temporal encounter network. (A) Degree distributions $P\left(k\right)$ of population and their neighbors (friends). The average degrees are ${{\left\langle k \right\rangle }_{control}}=238.5$ and ${{\left\langle k \right\rangle }_{sensor}}=442.0$, respectively. (B) Probability density function $P\left( f \right)$ of stage frequency of population and neighbor set. The inset shows the same plot in semi-log scale. The mean values are ${{\left\langle f \right\rangle }_{control}}=8.0$ and ${{\left\langle f \right\rangle }_{sensor}}=13.0$. (C) Probability density functions $P\left( {{k}_{s}} \right)$ of shell index ${{k}_{s}}$. The mean values are ${{\left\langle {{k}_{s}} \right\rangle }_{control}}=120.5$ and ${{\left\langle {{k}_{s}} \right\rangle }_{sensor}}=167.3$. (D) Distribution of encounter entropy $S$. The density function $P\left(S\right)$ has centralized peaks around $\ln 1=0$, $\ln 2=0.693$, $\ln 3=1.099$ and $\ln 4=1.386$, resulting from individuals with homogenous encounters in corresponding number of intervals. The mean values of encounter entropy are ${{\left\langle S \right\rangle }_{control}}=1.35\text{nat}$ and ${{\left\langle S \right\rangle }_{sensor}}=2.00\text{nat}$. The purple dashed lines in all these panels (from left to right) indicate the ${{90}^{th}}$, ${{99}^{th}}$, ${{99.9}^{th}}$ and ${{99.99}^{th}}$ percentiles of corresponding values across the whole population, explaining the degree of heterogeneity among most centrally located individuals, friend sensors and the population as a whole.}
\label{fig2}
\end{center}
\end{figure}

Considering that friend sensors are identified locally without using any centrality measures, they could be representative of a universally applicable strategy when it is costly or impossible to map the detailed network structure. To investigate the superiority of friend sensors in a comparable manner, we employed different centrality measures to quantify an individual's importance based on the both network structure and individual travel behavior employing the following centrality measures (see Supplementary Note 2): (1) Degree $k$, measuring total number of contacts of each individual during the week, (2) Travel frequency $f$, frequency of taking public transit services, ($f$ could also be interpreted as number of activities in temporal networks \cite{Perra2012}) (3) Shell index $k_s$, taken from $k$-shell decomposition \cite{Kitsak2010} on the static network and (4) Encounter entropy $S$, capturing temporal diversity of encounters:
\begin{equation}
S=-\sum\limits_{t} p_t \ln p_t,
\end{equation}
where $p_t$ is the probability of an individual's physical encounter in time $t$ (hourly). Using time-stamped encounter transactions, we can build the whole contact network and determine individual's centrality for both control and sensor sets (see Fig.~\ref{fig2}).

Indeed, a sensor group is more central than the randomly selected control group in terms of degree $k$ (Fig.~\ref{fig2}A); however, it is not yet known whether the friend paradox applies to other measures related to travel behavior (other than network structure). Before looking for additional sensors, we first measured other centrality distributions $P\left( f \right)$, $P\left( {{k}_{s}} \right)$ and $P\left( S \right)$ using both population and friend sensors. Although most people traveled less than 5 times during the week, we still found that $P\left( f \right)$ was characterized by a heavy tail across the population, indicating a significant degree of heterogeneity in individual transit use pattern (Fig.~\ref{fig2}B). Moreover, we found that $P\left( f \right)$ of the sensor group clearly exhibited the friend paradox as well, indicating that the people you have encountered on buses traveled more often than you do. Using the same control and sensor groups, we then obtained the distributions $P\left( {{k}_{s}} \right)$ and $P\left( S \right)$. As Fig.~\ref{fig2}C and D demonstrate, the friend paradox does exhibit in terms of shell index ${{k}_{s}}$ and encounter entropy $S$ as well, suggesting that friend sensors have higher $k$-shell indexes and show higher temporal encounter diversity than the population. Taken together, Fig.~\ref{fig2} suggests that the simple friend sensor scheme can universally identify more centrally located social sensors. Nevertheless, as the percentiles show (in all Fig.~\ref{fig2} panels), there are still significant differences between the most central individuals and friend sensors, further indicating that the efficiency of friend sensors might be limited. Taken together, as one might expect, the simple principle of friend sensor scheme also prevents itself from performing more efficiently, as better sensors could always be obtained by using more information on contact structure.

We next compare performance of the best sensors identified by each centrality measure against friend sensors by quantifying lead-time on a universal scale. When individual infected time cannot be obtained across the whole population, lead-time is estimated as difference between control and sensor samples in general \cite{Christakis2010}. However, since transit services are generally not operated 24 hours a day, the cumulative infection curve is not strictly monotonic increasing during the monitoring period in our case, resulting in significant difference when calculating lead-time from multiple runs; thus, using instantaneous lead-time is a biased measure of sensor performance (see Supplementary Note 3 and Supplementary Fig.~\ref{figs2}). However, given that individual infection time can be traced from simulations, we can essentially quantify lead-time against the whole population instead of a small sample control group. For efficient early detection, we fixed the monitored infected ratio $\hat{\alpha }=\left[ {{\alpha }_{1}},{{\alpha }_{2}} \right)=\left[ 0.05,0.25 \right)$ and measured only the difference of infection time of people in $\hat{\alpha}$, obtaining infection time $t_{P}^{{\hat{\alpha }}}=\left\{ {{t}_{i}}|{{\alpha }_{1}}\le {{F}_{P}}\left( {{t}_{i}} \right)<{{\alpha }_{2}} \right\}$ from population and $t_{S}^{{\hat{\alpha }}}=\left\{ {{t}_{i}}|{{\alpha }_{1}}\le {{F}_{S}}\left( {{t}_{i}} \right)<{{\alpha }_{2}} \right\}$ from sensor group ($F$ represents the empirical distribution of exposed time). We re-define lead-time as the difference of average $t_{P}^{{\hat{\alpha }}}$ and $t_{S}^{{\hat{\alpha }}}$:
\begin{equation}
  T = \left\langle t \right\rangle _P^{\hat \alpha } - \left\langle t \right\rangle _S^{\hat \alpha }
\end{equation}

\begin{figure}[tbp]
\begin{center}
\centerline{\includegraphics[scale=0.65]{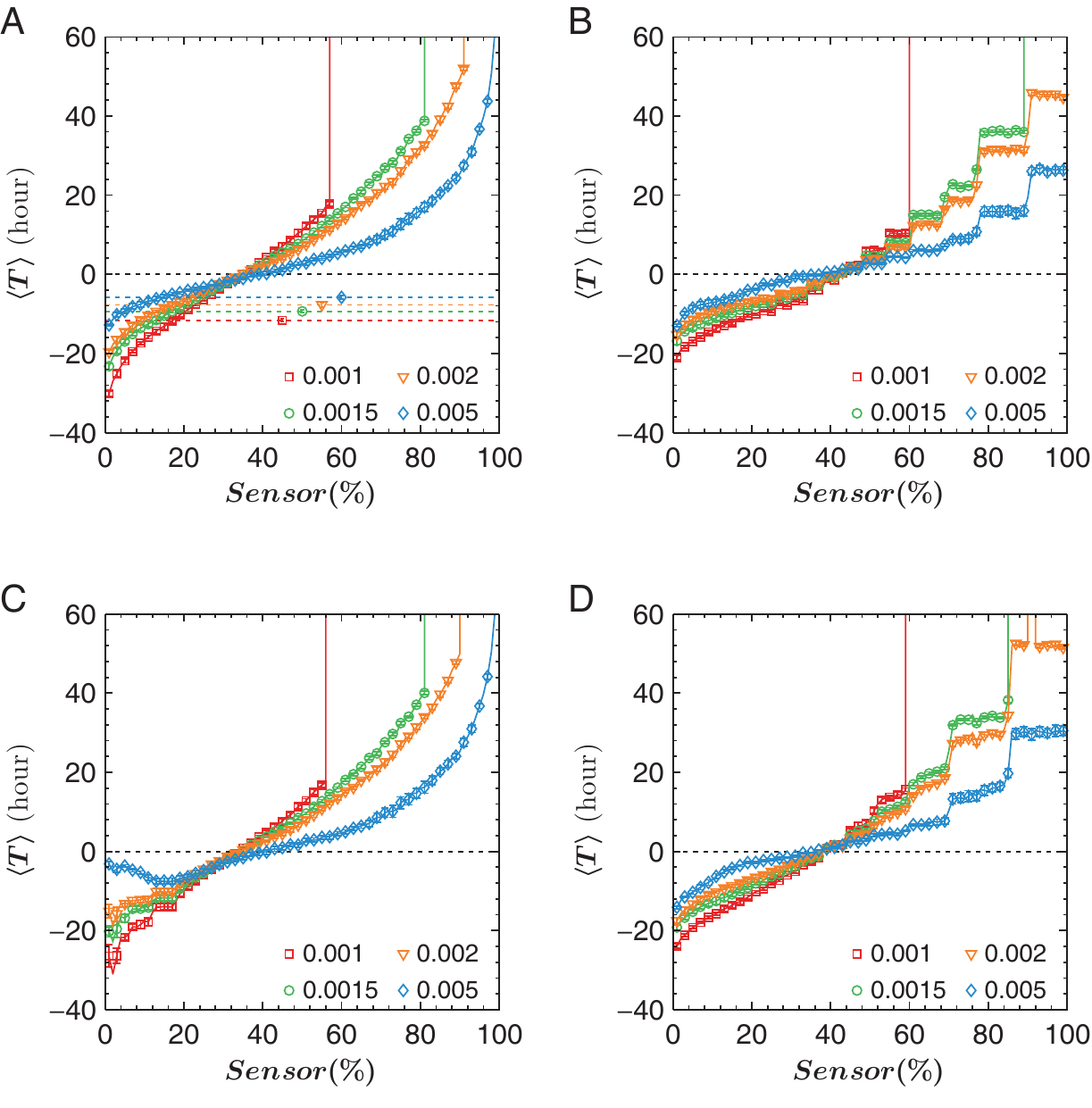}}
\caption{(A)-(D). Mean and standard deviation of lead-time for sorted slices (1\%) obtained by (A) degree $k$, (B) frequency $f$, (C) $k$-shell index ${{k}_{s}}$ and (D) encounter entropy $S$. In panel (A), the dashed line and error bars show lead-time provided by 1\% friend sensors as a guide. As no centrality measure is used in identifying friend sensors, lead-time will not change by choosing alternative control groups. All curves demonstrate a monotone increase approximately - except sensors identified by $k_s$; the top 1\% even fall behind friend sensors when $\beta=0.005$.}
\label{fig3}
\end{center}
\end{figure}

Next, we ordered individuals according to their centrality measure and divided the whole population into 100 percentiles. Using each percentile as a sensor group, we performed 20 simulation runs and measured the corresponding lead-times under different contagion rate $\beta $. As Fig.~\ref{fig3} shows, the top 1\% slices from all these partitions are able to provide early detection; however, the less the average centrality is, the shorter the lead-time $T$ will be. For example, the sensor group provides no advanced detection when $k\approx {{k}_{0.4}}$ and even falls behind the general population when $k>{{k}_{0.4}}$ (${{k}_{0.4}}$ is the ${{40}^{th}}$ percentile of degree). In this case, lead-time may reach infinity if the spreading cannot reach ${{\alpha }_{2}}$ (25\%) among sensor group. By comparing these centrality measures jointly, we found that they actually vary consistently on sensor composition; however, no one outperforms the others significantly (see Supplementary Fig.~\ref{figs2}).

\begin{figure}[tbp]
\begin{center}
\centerline{\includegraphics[scale=0.65]{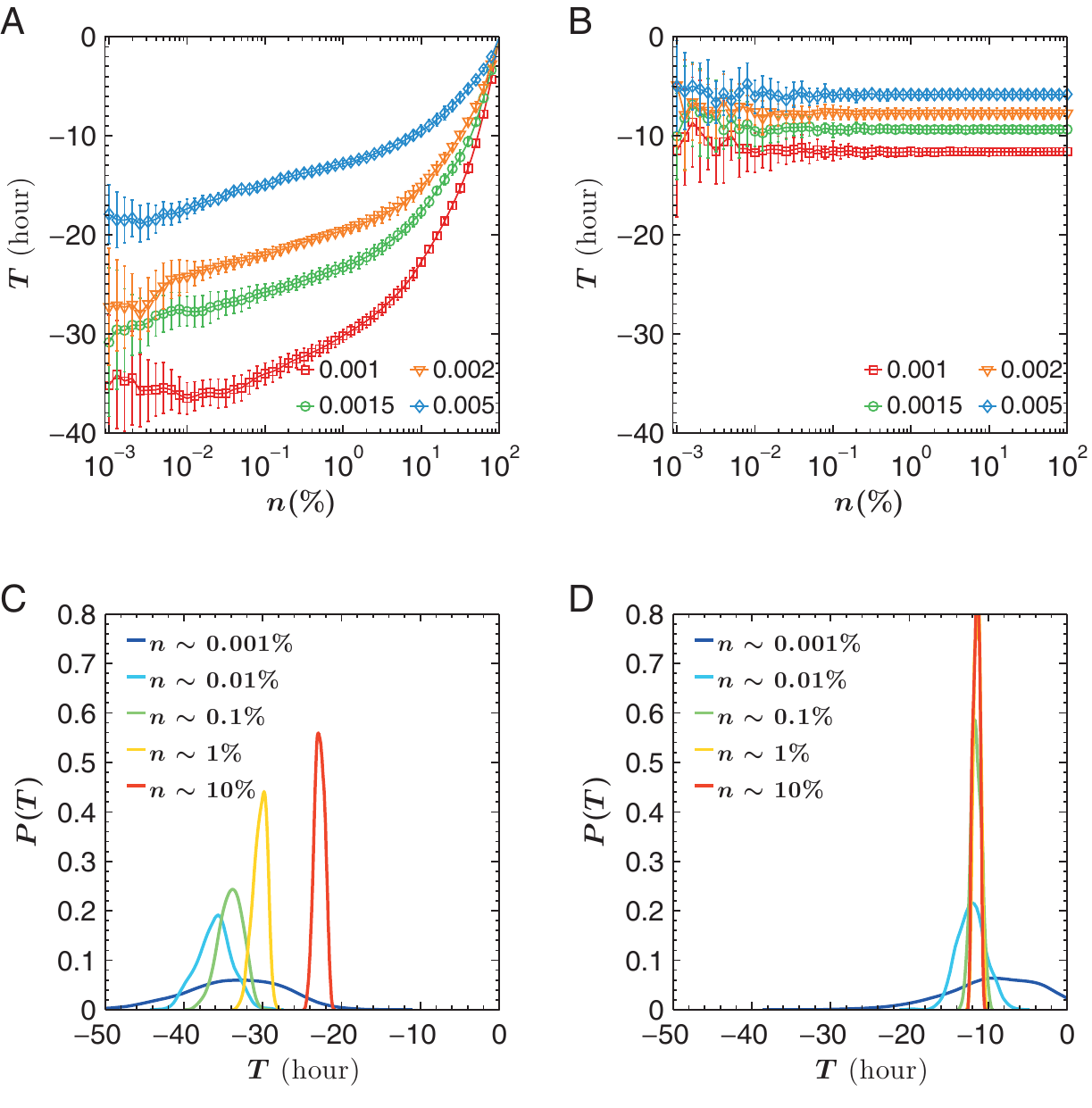}}
\caption{Effect of sensor size on efficiency and reliability in detecting contagious outbreaks. (A) Lead-time provided by sensors with highest degree, with sampling rate $n=\left| S \right|/\left| P \right|$ in a logarithmically spaced interval spanning from 0.001\% to 100\% with different contagion rate $\beta =\left\{ 0.001,0.0015,0.002,0.005 \right\}$. The error bars correspond to standard deviation of $T$. (B) Lead-time provided by friend sensors identified by random control group $C$ of different size. Given that sensors are characterized by the same distribution, lead-time exhibits a convergence pattern with the increase of sample size. In fact, with sampling rate $n$ increases, the variance of $T$ determined from one particular simulation run reduces, resulting in the decreasing overall variance. (C) Distribution $P\left(T\right)$ of lead-time $T$ given different sensor size $\left| S \right|$ when contagion rate $\beta =0.001$, corresponding to panel (A). (D) The same plot as panel C, however, for friend sensors corresponding to panel (B).}
\label{fig4}
\end{center}
\end{figure}

The efficiency of using such sensors to detect contagious outbreaks depends not only on centrality measures, but also sensor size $\left| S \right|$. On one hand, a small sample size induces large variation, providing poor reliability in potential applications. On the other hand, the difference of average centrality measure might be more and more significant given the intrinsic heterogeneity of individual behavior, revealing that we may achieve longer lead-time with lower cost (if the cost is in proportion to sensor size). In Fig.~\ref{fig4}A, we chose degree as primary centrality and measured lead-time for logarithmically spaced sampling rate $n=\left| S \right|/\left| P \right|$, spanning from 0.001\% (only 27 people with highest degree) to 100\% (the full population is used as sensors; lead-time is zero in this case). As the figure shows, smaller sample size indeed provides longer lead-time, but, with larger variation. In Fig.~\ref{fig4}B, we show performance of friend sensors obtained from equally sized control groups. Given that the sensor group is always sampled from a deterministic population, we observed a constant average lead-time, independent of sampling rate $n$. However, the standard deviation of lead-time decreases as sample size gets larger in both Fig.~\ref{fig4}A and B, corresponding to the law of large numbers when calculating lead-time in each simulation.

In practice, one should not just consider average lead-time and monitoring cost of such sensors; their reliability is equally important. To evaluate sensor reliability, we created a simulation result set with 500 runs and measured the lead-time distribution $P\left(T\right)$ for contagion rate $\beta=0.001$. As Fig.~\ref{fig4}C shows, average lead-time of different sensor groups (in terms of sensor sizes) is well characterized by normal distribution, however, with significant mean and variance difference. Notably, the top 0.01\% group performs extremely well for both average lead-time provided and reliability. Fig.~\ref{fig4}D shows results of the same analysis for the friend sensor scheme. We observed that the larger the sensor size is, the more reliable the lead-time becomes; however, increasing sensor size does not raise average performance, consistent with what Fig.~\ref{fig4}B shows. We also applied this procedure to other centrality measures: frequency $f$, $k$-shell index $k_s$ and encounter entropy $S$, finding that sensor group identified by degree outperforms all other centrality measures (see Supplementary Note 4 and Supplementary Fig.~\ref{figs4}). Taken together, Fig.~\ref{fig4} suggests that the friend sensor scheme indeed provides a substantial lead-time in early detection; however, the inherent principle prevents it from performing better by adjusting sensor sizes (in other words, average performance is independent on sensor size), whereas a well-defined sensor (obtained by degree centrality in this case) can easily outperform it. Our results further illustrate a clear advantage of deriving sensors from the full co-presence network, providing longer, more reliable lead-time by using a smaller sensor group.

\section{Discussion}

To summarize, we show the feasibility of a friend sensor scheme in providing early detection during a contagious outbreak in a metropolitan physical contact network. Indeed, the simple friend sensor scheme, which does not require a detailed network structure, works consistently well in finding sensors that are more central in the network. However, since all friend sensors are actually characterized by a deterministic neighbor population based on network structure, their performance is often limited by inherent characteristics of the neighbor population, providing constant early warning on average, independent of sample selection and sample size. Therefore, it is still crucial to show the value of full network structure, in particular for early detecting contagious outbreaks. Taking advantage of individual-based passive data collection techniques on city-scale (transit fare collection systems in this paper), we mapped detailed spatial and temporal structures as a whole and identified new sensors given diverse centrality measures, offering new insight into finding more efficient social sensors. Considering the weak, passive and indirect nature of social links enabled by these common daily physical encounters, $k$-shell index $k_s$ - a well-defined structural centrality - is less effective than the simple degree $k$ and frequency $f$ (number of activities) in contagious detection. Note that we did not use betweenness and closeness centralities as a measure in our study. On one hand, computing shortest path-based centralities is extremely time-consuming because of this network's high density. On the other hand, considering the temporal nature of daily encounters, the role of static shortest path is not as significant as it is in social networks of personal relations. Based on the spreading settings examined in our study, a well-defined social sensor group based on degree may account for only 0.01\% of the total population; however, it provides longer and more reliable lead-time - than the friend sensor scheme - allowing public health officials and governments to plan a quick and efficient response. In practice, those sensor individuals can be easily identified by transit agencies given their unique smart card IDs, and the remaining question is to monitor the health status of sensor groups. Although sensor groups are deterministic given our observations, it may still not easy for public health agencies to monitor their status owing to privacy and technical issues, which are beyond the goal of our study. Nevertheless, one possibility is to use the emerging ICT in health monitoring - such as health-care applications in smartphones - and ask these special people to willingly provide anonymous information to public health agencies. Another possibility, assuming that health authorities could have access to the identities of these ¡°sensor group¡± individuals, is to, instead of monitoring them directly, track their appearance as cases of selected contagious diseases reported by hospitals and local clinics. Were this sensor groups to appear with a certain statistical trend in clinical reports, we would have in our hands an early-warning signal for the future advancement of the contagious disease.

Having said that, influenza-like diseases are transmitted primarily by close contacts. Although the network used in our study is created across the whole metropolitan area, capturing all transit users' contacts during a whole week, it still covers only a small slice of all potential contacts in our daily life, forming only a subnetwork of a network of contacts that would be important in the spread of actual epidemics. On the other hand, to simulate an outbreak, we fixed relatively unrealistic simulation settings, such as introducing only 2-hour exposed period and using an simplified SEI model instead of a full developed SIR or SEIR model, for the outbreak to travel at a speed where global and local methodologies could be tested. To what extent the simulation can match a real contagious outbreak and the relevance of the simulations findings to actual epidemics remain to be measured. Thus, it is important to note that the specific results in our study are embedded in the physical encounter network with a pre-defined spreading mechanism. Such encounters on transit vehicles occur more often between perfect strangers than among friends, colleagues or families, making the network incomplete for predicting epidemic spreading via all possible transmission pathways. Therefore, great caution should be exercised in interpreting the results. In reality, a full contact network for disease spreading consists of all of social links from diverse circumstances; it remains unclear to us which part should be given priority with respect to the characteristics of an unknown virus/disease. Nevertheless, with the rapid development of information and communication technologies, mapping the whole structure of close encounters from various data would be far less difficult and laborious today. Given the high individual and collective regularities rooted in human behaviors \cite{deMontjoye2013,Gonzalez2008,Song2010,Sun2013}, patterns of face-to-face encounters in various settings could be documented as well \cite{Salathe2010,Stehle2011}, helping us build more comprehensive agent-based models to contain emerging epidemics \cite{Eubank2004,Smieszek2011}. Moreover, with our increasing knowledge about ourselves and various microorganisms around us, more efficient social sensors for different scenarios can be identified and applied in monitoring contagious spreading from day to day, providing early and accurate information to support better decision making. We believe that our work can serve as a base to help better combat the spread of disease on a citywide scale \cite{Ancel2003,Ferguson2005} and better understand social contagion dynamics \cite{Christakis2007,Bond2012,Christakis2013}.

\section{Methods}

\subsection{Data sets}

Trip records were collected from Singapore's smart-card-based fare collection system, covering more than 96\% of public transit trips. The system collects data for both bus and MRT (subway) modes. Smart card data is widely used in public transit: network planning, service adjustments, providing ridership statistics, and indicating service performance. We employ bus, not MRT (Mass Rapid Transit, railway based) trip records in this study, since it is difficult to identify close proximity interactions on large MRT trains. For buses, once a smart card holder boards a vehicle (tapping-in), the system generates a temporary transaction record; after he/she leaves the vehicle (tapping-out), a complete record will be stored with detailed trip information.

A full bus trip may contain more than one stages with transfers from one route/vehicle to another. The stage records are generated separately in the smart card system (with each tapping-in and tapping-off). Since our goal is to identify in-vehicle encounters and the people one may encounter in vehicles will differ from stage to stage, we use the term trip to represent stage in this document. After processing the raw data, we obtained the trip records used in this study. The fields and their contents are provided in Supplementary Tab.~S1.

This study was performed on the trip records of one week in March, 2012. The dataset contains 22,455,159 bus trip transaction records from 2,969,320 individual smart card holders.

\subsection{Simulation}

To evaluate the performance of social sensors in the obtained interaction network, we use the SEI models to simulate contagious outbreaks among all transit users \cite{Anderson1992}, which are assumed to be in one of two states: susceptible (S) when they are prone to infection, exposed (E) between exposure and infectiousness, or infected (I) when they can transmit the disease to others. In studying the outbreak dynamics, we are more interested in the initial spreading processes and thus we do not consider the recovery stage in the simulation.

All simulations start on Saturday and end on the next Friday, spanning the whole week (given the dataset). In the spreading process, the duration of explosive stage (such that $0<I/N<1\%$) is highly determined by the number of index cases. Thus, a smaller index size induces larger in terms of temporal spreading processes; however, after this explosion, the spreading becomes steady and contagion rate $\beta$ determines the spreading speed of the rest spreading. Thus, to boost the initial spreading processes, we set ten index cases in our simulation, enabling us to observe outbreaks in one week. On the other hand, since people show great heterogeneity in their transit use behavior (such that $f\le 5$ for almost 50\% of the users during the week), a larger number of index cases also prevents the disease from dying out at initial stage. However, as sensor performance is monitored given infected ratio $5\%\le F_P(t_i) <25\%$ (after the explosive stage and during the steady spreading), lead-times and their variability are mainly determined by $\beta$ rather than number of index cases (see Supplementary Fig.~\ref{figs5}). Thus, all our simulations start with ten infected people (ten infected cases), randomly selected across all transit users who were active on Saturday (who took buses on Saturday). After being infectious, individual $i$ will transmit disease to a susceptible individual $j$, who individual $i$ encountered during his/her journey, with probability $p_{ij}=\beta\times d_{ij}$ ($d_{ij}$ is encounter duration). Here, $\beta$ is an important parameter determining the speed of contagious spreading. We chose a series of values from 0.001 to 0.005 per 20 seconds. On one hand, these values are similar to the value used in Ref.~\cite{Salathe2010}. On the other hand, by simulating the spreading processes with different $\beta$, we can better evaluate the performance of different sensors for outbreaks with different $\beta$. Given any instantaneous network in a vehicle is a fully connected one, disease may spread very fast once one individual get infect. To avoid this, we introduce the exposure (E) stage - which lasts for 2 hours - in our simulation: once an individual is exposed, he/she will not spread the disease immediately; however, he/she begins to begin to spread the disease to other encounter people after 2 hours. Considering that most transit trips take place in under 2 hours, one is unlikely to get infected and begin spreading disease to others during the same trip.

In the simulation, each time step represents 0.5h, such as 7:00-7:30. In any step $t$, we first identify all the neighbors he/she has encountered (the time they encountered each other should be within this time step) and then get them exposed with the defined probability $\beta$. The incubation time is selected as a constant given the time granularity (2h), and thus, the exposed individuals become infectious in step $t+4$. We also tested our results when setting the exposed period to be 6h and 12h, finding that sensors identified by degrees performs consistently better than others (Supplementary Fig.~\ref{figs6}).

Based on these simulation settings, one can monitor the temporal spreading dynamics from a set of simulations with certain $\beta$ and random seeds as initial infected people. Meanwhile, individual infection time could be traced from each simulation. As Supplementary Fig.~\ref{figs1} shows, although contagion rate $\beta$ in each panel is the same, simulations still differ significantly from each other, in particular when $\beta$ is low. Thus, estimating lead-time universally is important to establish the difference.

\section{Acknowledgments}
We thank Singapore's Land Transport Authority for providing the smart card data. Special thanks to the developers of igraph library. This work was supported by National Research Foundation of Singapore, which is the funding authority of Future Cities Laboratory. Manuel Cebrian is funded by the Australian Government as represented by the Department of Broadband, Communications and Digital Economy and the Australian Research Council through the ICT Centre of Excellence program.

\subsection{How to cite this article}
{Sun, L., Axhausen, K.W., Lee, D.-H., Cebrian, M., Efficient detection of contagious outbreaks in massive metropolitan encounter networks. \newblock {\em Scientifit Reports} {\bf 4}, 5099 (2014). DOI: 10.1038/srep05099.}

\appendix
\renewcommand{\thefigure}{S\arabic{figure}}
\renewcommand{\thetable}{S\arabic{table}}
\renewcommand{\theequation}{S\arabic{equation}}
\setcounter{figure}{0}
\setcounter{table}{0}
\setcounter{equation}{0}

\newpage

\section{SUPPORTING INFORMATION}
\section{SUPPLEMENTARY TABLE S1}
\begin{table}[htbp]
\caption{Fields and contents of trip record dataset}
\begin{tabular}{lp{5.5cm}}
\hline
    Field & Description \\
    \hline
    Trip ID & A unique number for each transit trip\\
    Card ID & A unique coded number for each smart card (anonymised) \\
    Passenger Type & The attribute of cardholder (Adult, Senior citizen and Child) \\
    Service Number & Bus route service number (e.g. 96) \\
    Direction & Direction of the bus route (0 and 1) \\
    Bus Registration No. & A unique registration number for each vehicle (e.g. `0999') \\
    Boarding Stop ID & A unique number for boarding stop (e.g. 40009) \\
    Alighting Stop ID & A unique number for alighting stop (e.g. 40009) \\
    Ride Date & Date of a trip (e.g. `2012-03-26') \\
    Ride Start Time & Start (tapping-in) time of a trip (e.g. ¡¯08:00:00¡¯) \\
    Ride End Time & End (tapping-out) time of a trip (e.g. ¡¯08:00:00¡¯) \\
    Ride Distance & Distance of the trip (e.g. 12.0 km) \\
\hline
\end{tabular}
\label{tabs1}
\end{table}

\section{SUPPLEMENTARY NOTE 1: ENCOUNTER NETWORK}

The physical network is built by identifying all in-vehicle encounters (two individuals occupying the same bus at the same time) using smart card transaction records. Based on the detailed bus registration number and boarding/alighting time of each bus trip, one can extract all encounters on a particular vehicle by checking whether any two bus trips overlap in time. By applying this procedure for the whole week, we obtained a temporal contact network with 3 million vertices and 1 billion edges (with all transit users as vertices and physical encounters between them as edges), across all of metropolitan Singapore. Vertex attributes include Card ID and Passenger Type; edge attributes contain encounter time and encounter duration. The duration of each encounter is used to model the weight of each social link. Detailed structural characteristics of an encounter network can be found in Ref.\cite{Sun2013}.

In fact, many of the short encounters (less than 1 minute) are created by passengers' simultaneous boarding and alighting at a particular bus stop without incurring any physical contacts. Considering vehicle configuration and load profile, occupying the same vehicle does not necessarily imply an intense social contact, i.e. talking to each other. To account for the effect of these short encounters which result in no real interactions, we removed all edges with $d_{ij}< 5 \,\text{min}$ when simulating the contagious outbreak. The final network for simulation consists of 2.7 million vertices and 0.3 billion edges.

\section{SUPPLEMENTARY NOTE 2: CENTRALITY MEASURES}
\label{notes2}
\subsection{Degree}

Degree of vertex $i$ is defined as the number of neighbors (contacts in this article):
\begin{equation}
  k\left(i\right) = \sum\limits_{j\in N\left(i\right)} a_{ij},
\end{equation}
where $N\left(i\right)$ is the neighbor set of vertex $i$ and $a_{ij}=0,1,2,\cdots$ is the number of edges (contacts) between $i$ and $j$. Degree is a local index without considering the importance of neighbors.

\subsection{Frequency}

Given the strong heterogeneity exhibited in individual transit use pattern, travel frequency $f(i)$, which register the number of times individual $i$ took public transit services during the studied week, could be a measure of centrality as well. This measure could be interpreted as number of activities in temporal interactions \cite{Perra2012}.

\subsection{$k$-shell Index}

The $k$-shell index $k_s$ is obtained from $k$-shell (or $k$-core) decomposition \cite{Kitsak2010}. The decomposition process starts with removing all vertices with $k=1$ recursively until $k\ge 2$ for all the remaining vertices, assigning removed vertices with $k_s=1$. By increasing degree to $k=2$, we can continue the process, finding vertices with $k_s=2$. In the same manner, all vertices in the network can be separated into groups with different $k_s$ value, which is called the $k$-shell index.

$k$-shell index incorporates an individual vertex's location, providing a better measure to quantify individual importance in spreading processes. However, its performance is not good in spreadings with multiple index cases, since the vertices with large $k_s$ are usually clustered with each other while those with high degrees tend to be distributed uniformly across the population.

\subsection{Encounter Entropy}

As social encounters in daily life vary significantly with time, the importance of one vertex in temporal spreading processes depends on the temporal diversity of individual's travel behavior. Thus, we define encounter entropy $S$ - as a special centrality for this study  - to measure the diversity of one's travel during the week, as:
\begin{equation}
  S\left(i\right) = -\sum\limits_{t\in T} p_t \ln p_t,
\end{equation}
where $p_t$ is the probability that individual $i$ encounters others at time $t$.

\subsection{Eigenvector Centrality}

Eigenvector centrality $C_e\left(i\right)$ of vertex $i$ is defined as
\begin{equation}
  C_e\left(i\right) = \lambda^{-1}\sum\limits_{j=1}^N a_{ij}e_{ij},
\end{equation}
where $\lambda$ is the maximum eigenvalue of adjacent matrix $A=\left[a_{ij}\right]$ and $e = \left(e_1,e_2,\cdots,e_n\right)^{T}$ is the eigenvector corresponding to $\lambda$.

\subsection{Closeness Centrality}
Closeness centrality $C_c(i)$ measures the impact of vertex $i$ on other vertices across the whole network, defined as:
\begin{equation}
  C_c\left(i\right) =\frac{N-1}{\sum\limits_{j=1}^N d_{ij}},
\end{equation}
where $d_{ij}$ is length of shortest-path between $i$ and $j$.
\subsection{Betweenness Centrality}
An important measure in social networks, betweenness centrality $C_b(i)$ is defined as:
\begin{equation}
  C_b\left(i\right) = \sum\limits_{s<t} \frac{n_{st}^i}{g_{st}},
\end{equation}
where $g_{st}$ is the number of shortest-paths between vertex $s$ and $t$ and $n_{st}^i$ is number of paths in $g_{st}$, which includes vertex $i$. The higher $C_b(i)$ is, the more influential vertex $i$ is in the network. In a weighted network, the shortest path is usually calculated using inverse weights.

When comparing sensor performance based on different centrality measures, we did not take those with shortest path-based centrality measures into consideration. On one hand, as social encounters happen frequently among various subjects, the contact network in our study has a very high density and computing these measures could be very costly. On the other hand, social encounters are highly determined by time; however, the shortest path computation on a static network is independent on temporal information.

\section{SUPPLEMENTARY NOTE 3: LEAD-TIME}\label{notes3}
Lead-time is a crucial measure in evaluating sensor performance from simulation results. Previous attempts quantify lead-time as the difference of infection time between control/sensor groups. However, in a simulation, we can actually trace the infection times of all individuals to better define the cumulative infection curve $\alpha(t)=I(t)/N(t)$. As Supplementary Fig.~\ref{figs2}A and B show, lead-time actually varies with both $\beta$ and $\alpha(t)={I(t)}/{N(t)}$. More importantly, it also varies from one simulation to another. Therefore, it might be biased to use lead-time for a fixed $\alpha$ to access sensor performance (see Supplementary Fig.~\ref{figs2}C and D), especially when there is a strong degree of heterogeneity for different simulation runs.

To better quantify a sensor's importance in detecting outbreaks early, we define the observation interval of $\alpha$ as $\hat{\alpha}=\left[0.05,0.25\right)$. We chose these two values since they might be representative for lower and upper bound to characterize the infection ratio for a contagious outbreak.

\section{SUPPLEMENTARY NOTE 4: SENSOR COMPARISION}
\label{notes4}
To access performance of different centrality measures jointly, we measured the average trend of centrality change in Supplementary Fig.~\ref{figs3} in terms of population sliced obtained by corresponding centrality measures. For example, in Supplementary Fig.~\ref{figs3}, we show the change of average degree $\langle k\rangle$ of the 100 slices obtained by travel frequency (number of activities) $f$, $k$-shell index $k_s$ and encounter entropy $S$, respectively. In this case, the sensors identified with maximal travel frequency, maximal $k$-shell index and maximal encounter entropy exhibit an average degree of $\langle k\rangle _f =918.2$, $\langle k\rangle _{k_s} =899.5$ and $\langle k\rangle _S =1069.0$, respectively.

As Supplementary Fig.~\ref{figs3} shows, all centrality measures are consistent with each other, displaying decreasing treads of average centrality on sequential sensor composition (each covering 1\% of the population). Taking Supplementary Fig.~\ref{figs3} as a whole, we found no one outperforms the others significantly among these centrality measures. However, the top 1\% sensors identified by degree and encounter entropy are more centrally located in terms of other $k_s$ and $f$ essentially. In contrast, the best sensors identified by $k_s$ performs poorly in terms of $k$, $f$ and $S$. In fact, considering the fact that this interaction network is composed of common daily encounters -- social links which are weak, passive and indirect -- the structural (global) centrality $k$-shell index
may not carry as much information as it would in a social network enabled by personal relations. On the contrary, given the small variation of $k_s$ compared to other local centrality such as $k$ and $f$ (see Fig.~\ref{fig2}), there are more individuals sharing the same $k$-shell index than other centrality measures \cite{Borge2012}, preventing us from identifying the most influential spreaders among them. For example, as Supplementary Fig.~\ref{figs4} shows, the top sensors obtained by $k$, $f$ and $S$ are excellent while sensors obtained by top $k_s$ perform poorly and unreliably.

\onecolumngrid
\newpage
\section{SUPPLEMENTARY FIGURE S1}
\begin{figure}[h]
\centering
\includegraphics[scale=0.65]{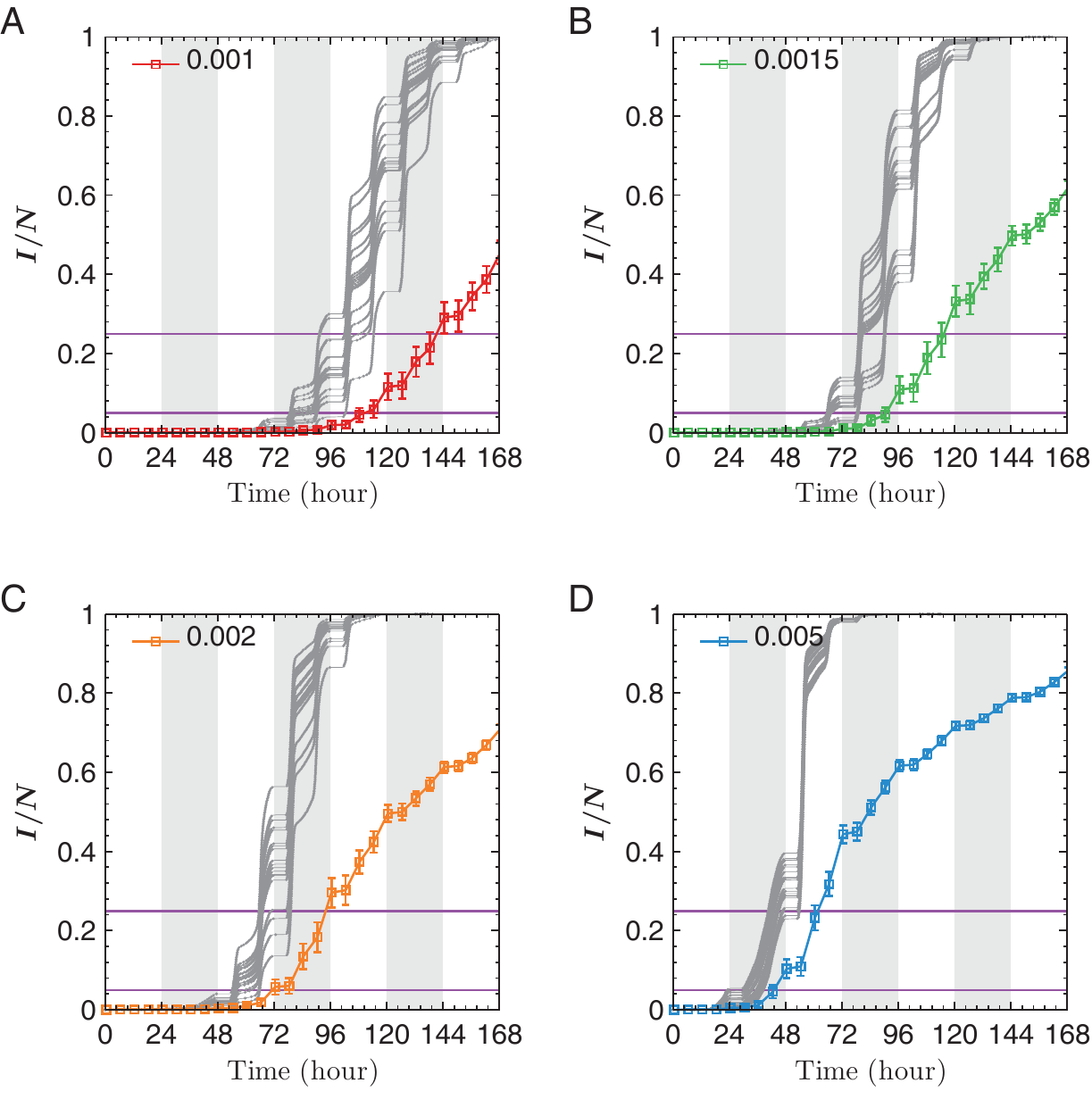}
\caption{The cumulative infected ratio $I/N$ over time for $0.01\%$ individuals with highest degree from 20 simulations for different contagious rate (A) $\beta=0.001$, (B) $\beta=0.0015$, (C) $\beta=0.002$, (D) $\beta=0.005$. In each panel, the temporal change of $I/N$ across the population is also shown for reference. The two solid lines correspond $I/N=5\%$ and $I/N=25\%$.}
\label{figs1}
\end{figure}

\newpage
\section{SUPPLEMENTARY FIGURE S2}
\begin{figure}[h]
\centering
\includegraphics[scale=0.65]{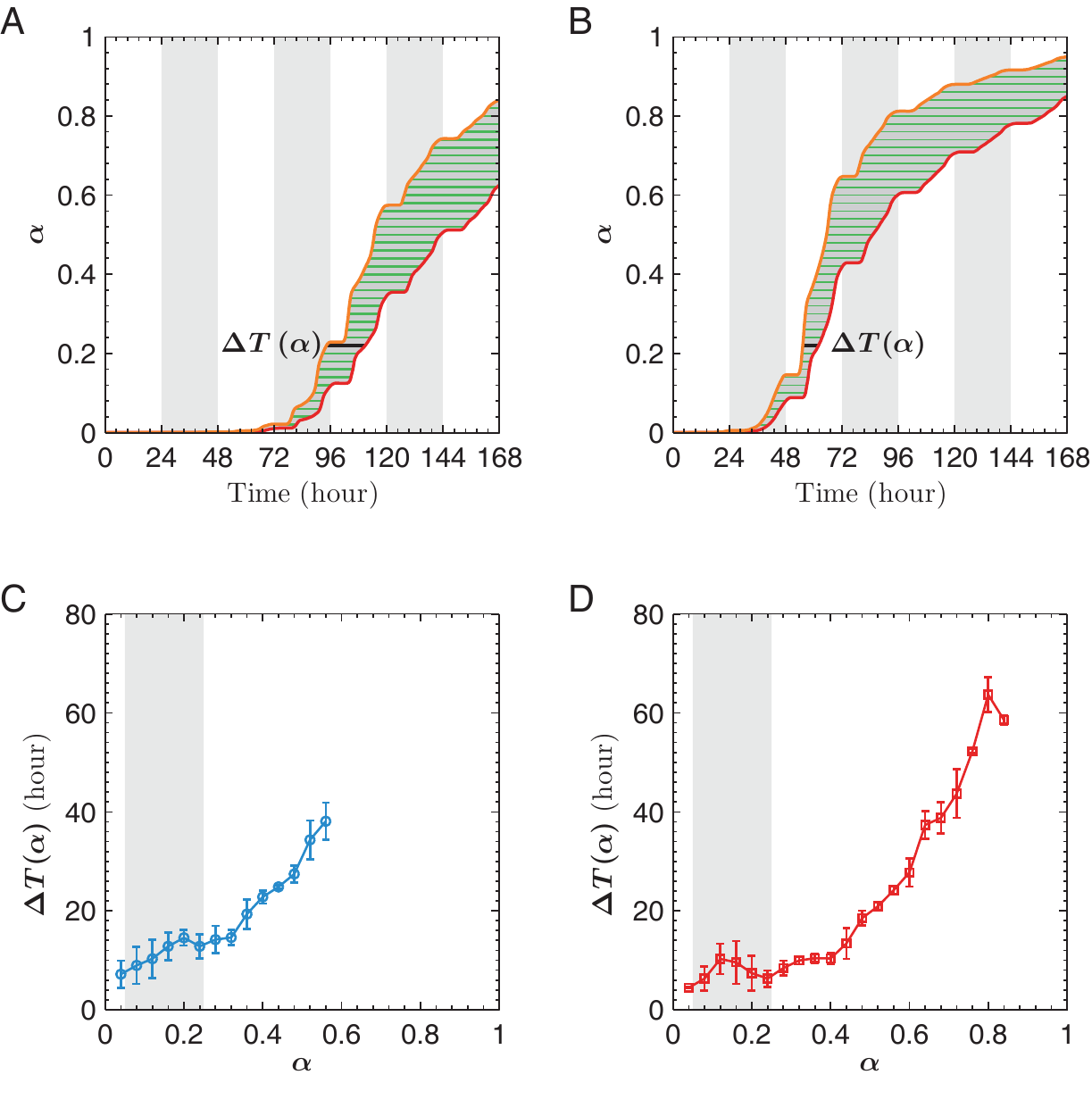}
\caption{Definition of lead-time. (A) The temporal change of infected ratio $\alpha (t)=I(t)/N(t)$ for $1\%$ randomly selected individuals (as control group; solid curves) and their neighbors (as sensors; dashed curves) with $\beta = 0.0015$. We can estimate lead-time $\Delta T(\alpha)=T_C(\alpha)-T_S(\alpha)$ given any infected ratio $\alpha$. (B) Same as in panel (A) but for $\beta = 0.005$. (C) Lead-time variation with $\alpha$ using the same control/sensor samples. The markers and error bars show the mean and standard deviation from 20 simulations. Given the temporal nature of transit activities, the spreading dynamics also varies with time and stops over night. Thus, the lead-time $\Delta T$ actually varies significantly with $\alpha$. (D) The same plot as in panel (C), but for $\beta=0.005$.}
\label{figs2}
\end{figure}

\newpage
\section{SUPPLEMENTARY FIGURE S3}
\begin{figure}[h]
\centering
\includegraphics[scale=0.65]{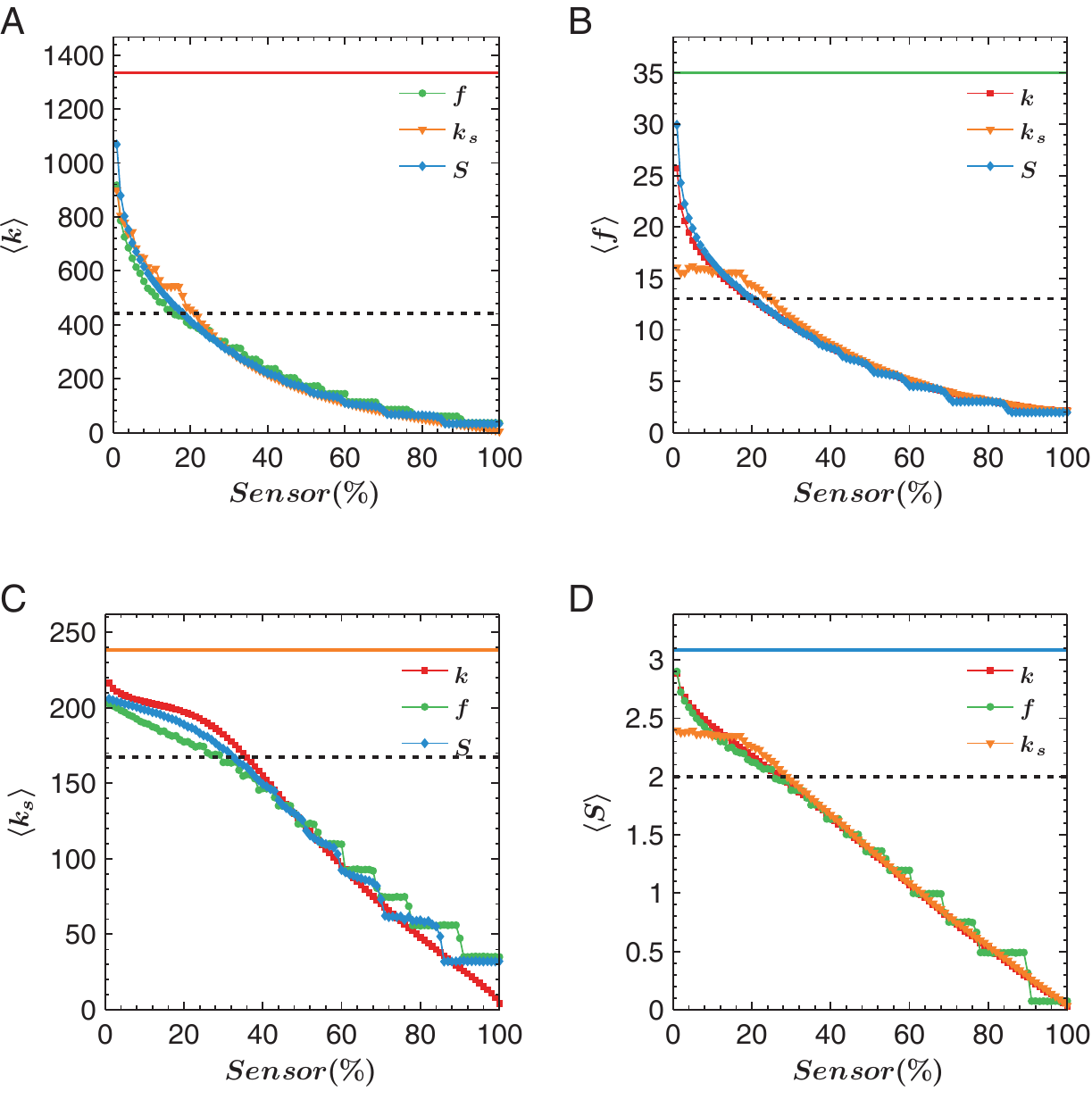}
\caption{Comparison of different centrality measures. (A) The average degree $\langle k\rangle$ for sorted slices (1\%; equal size) obtained by frequency $f$, $k$-shell index $k_s$ and encounter entropy $S$. The solid line and the dashed line show $\langle k\rangle$ of the top 1\% ($\langle k\rangle_k=1335.0$) and the whole population ($\langle k\rangle_P=442.0$), respectively. The average degrees of best sensors identified by other centrality measures are $\langle k\rangle_f=918.2$, $\langle k\rangle_{k_s}=899.5$ and $\langle k\rangle_S=1069.0$.
(B)-(D), same as panel (A), for other centrality measures. (B) The corresponding values for best sensors and population are $\langle f\rangle_f=35.0$, $\langle f\rangle_P=13.0$, $\langle f\rangle_k=25.7$, $\langle f\rangle_{k_s}=16.1$ and $\langle f\rangle_f=30.0$. (C) The corresponding values are $\langle k_s\rangle_{k_s}=238.3$, $\langle k_s\rangle_{P}=167.3$, $\langle k_s\rangle_{k}=216.5$, $\langle k_s\rangle_{f}=202.8$ and $\langle k_s\rangle_{S}=206.0$. (D) The corresponding values are $\langle S\rangle_{S}=3.09\text{nat}$, $\langle S\rangle_{k}=2.88\text{nat}$, $\langle S\rangle_{f}=2.90\text{nat}$ and $\langle S\rangle_{k_s}=2.40\text{nat}$.}
\label{figs3}
\end{figure}

\newpage
\section{SUPPLEMENTARY FIGURE S4}
\begin{figure}[h]
\centering
\includegraphics[scale=0.65]{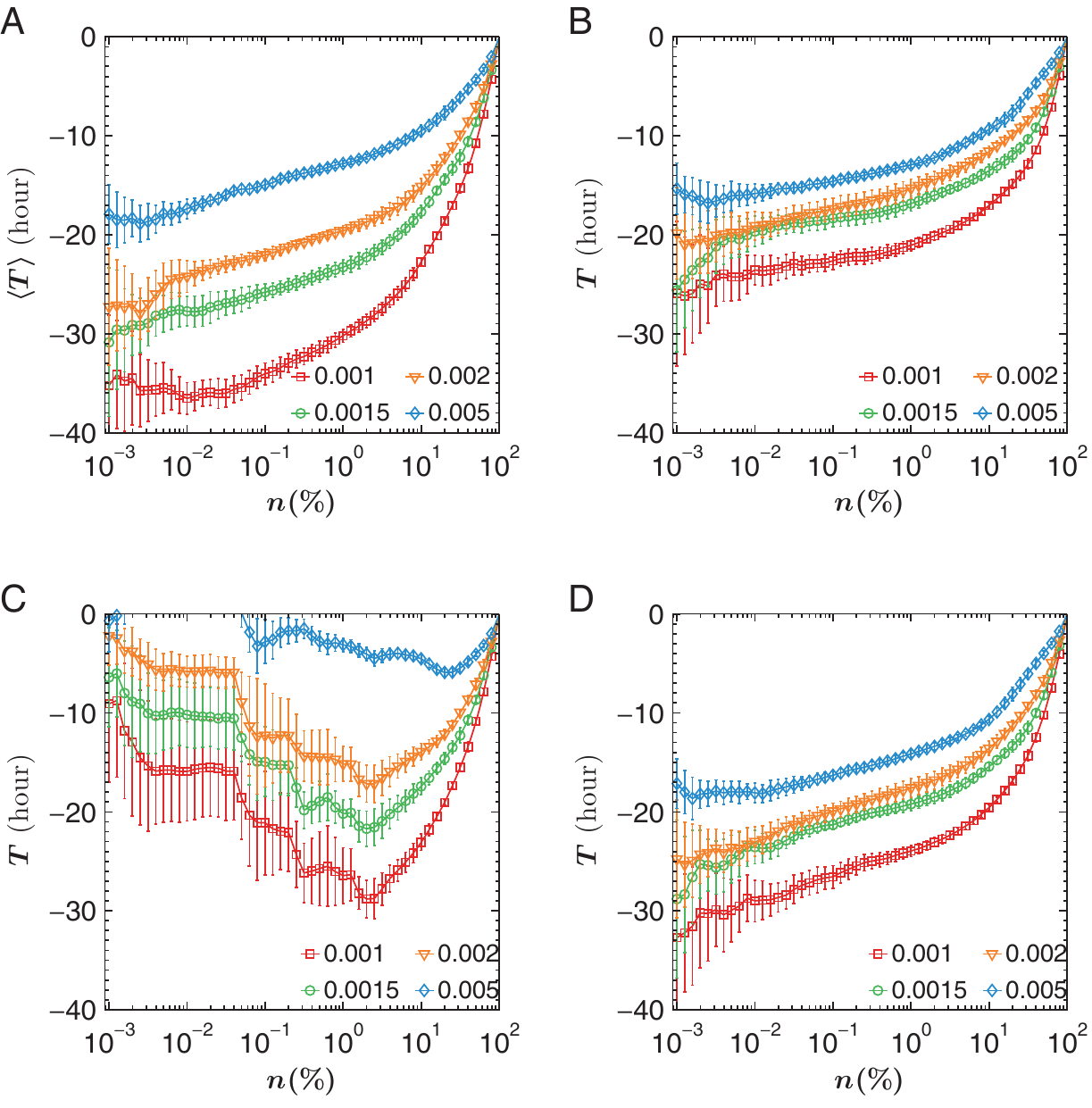}
\caption{Effect of sensor size on detection efficiency. (A) Lead-time provided by sensors with highest degree, with sampling rate $n=\left|S\right|/\left|P\right|$ in a logarithmically spaced interval spanning from 0.001\% to 100\% with different contagious rate $\beta=\{0.001,0.0015,0.002,0.005\}$. The error bars correspond to standard deviation of $T$. (This is the same plot as Fig. 4A.) (B)-(D), same plots as panel (A); however, for other centrality measures: (B) Frequency $f$; (C) $k$-shell index $k_s$ and (D) Encounter entropy $S$.}
\label{figs4}
\end{figure}

\newpage
\section{SUPPLEMENTARY FIGURE S5}
\begin{figure*}[h]
\centering
\includegraphics[scale=0.65]{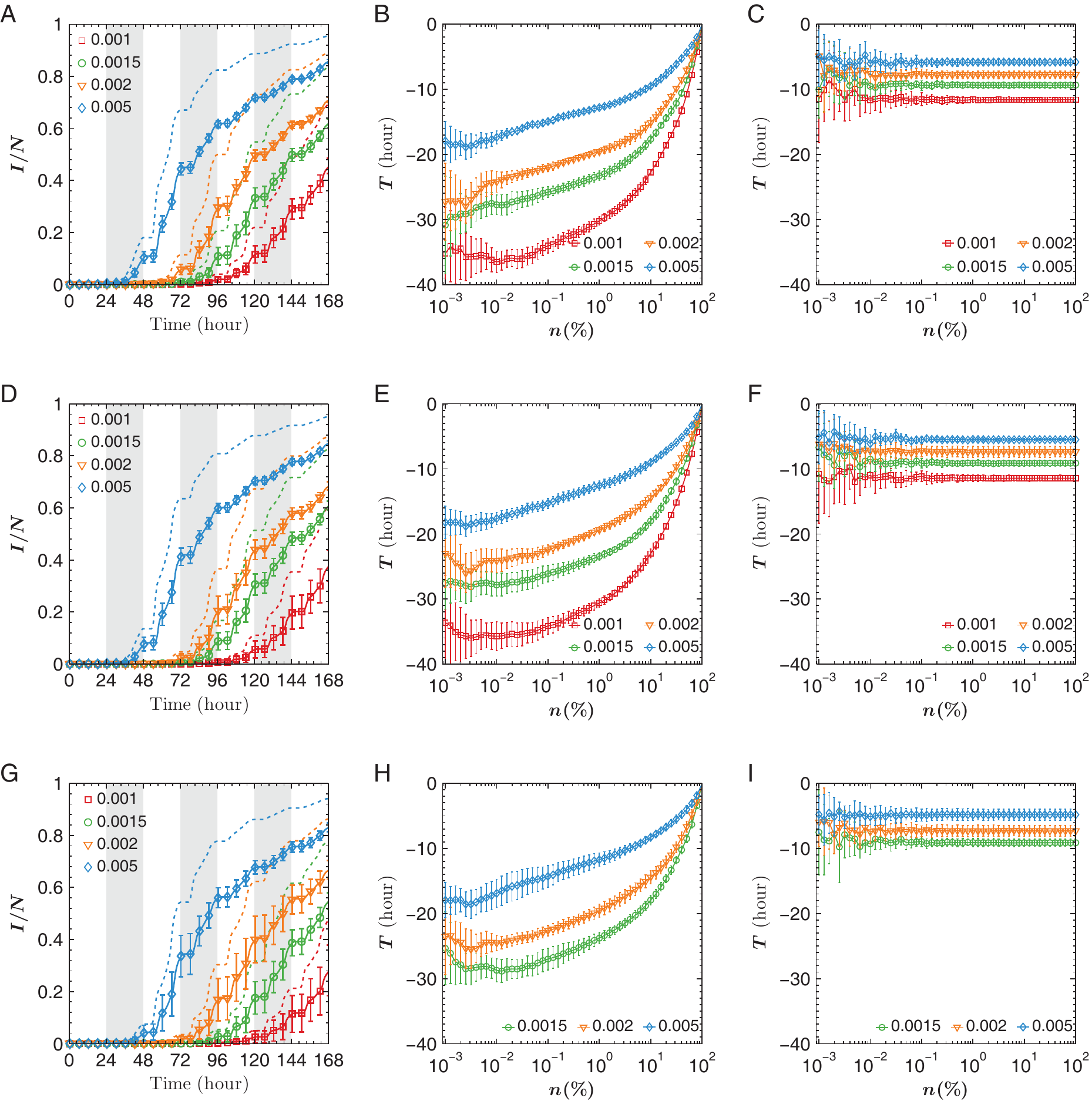}
\caption{Sensitivity on number of index cases (exposed for two hours). (A-C) Temporal ratio of infected cases (A), lead-time provided by sensors with highest degree (B) and friend sensors (C) when index case size is 10. (D-F) Index case size is 5. (G-I) Index case size is 2. In this case, the spreading cannot reach 25\% when $\beta=0.001$ after the whole period, so lead-time is not available. As lead-time is monitored after the explosive stage, $\beta$ determines the final lead-time observations.}
\label{figs5}
\end{figure*}

\newpage
\section{SUPPLEMENTARY FIGURE S6}
\begin{figure*}[h]
\centering
\includegraphics[scale=0.65]{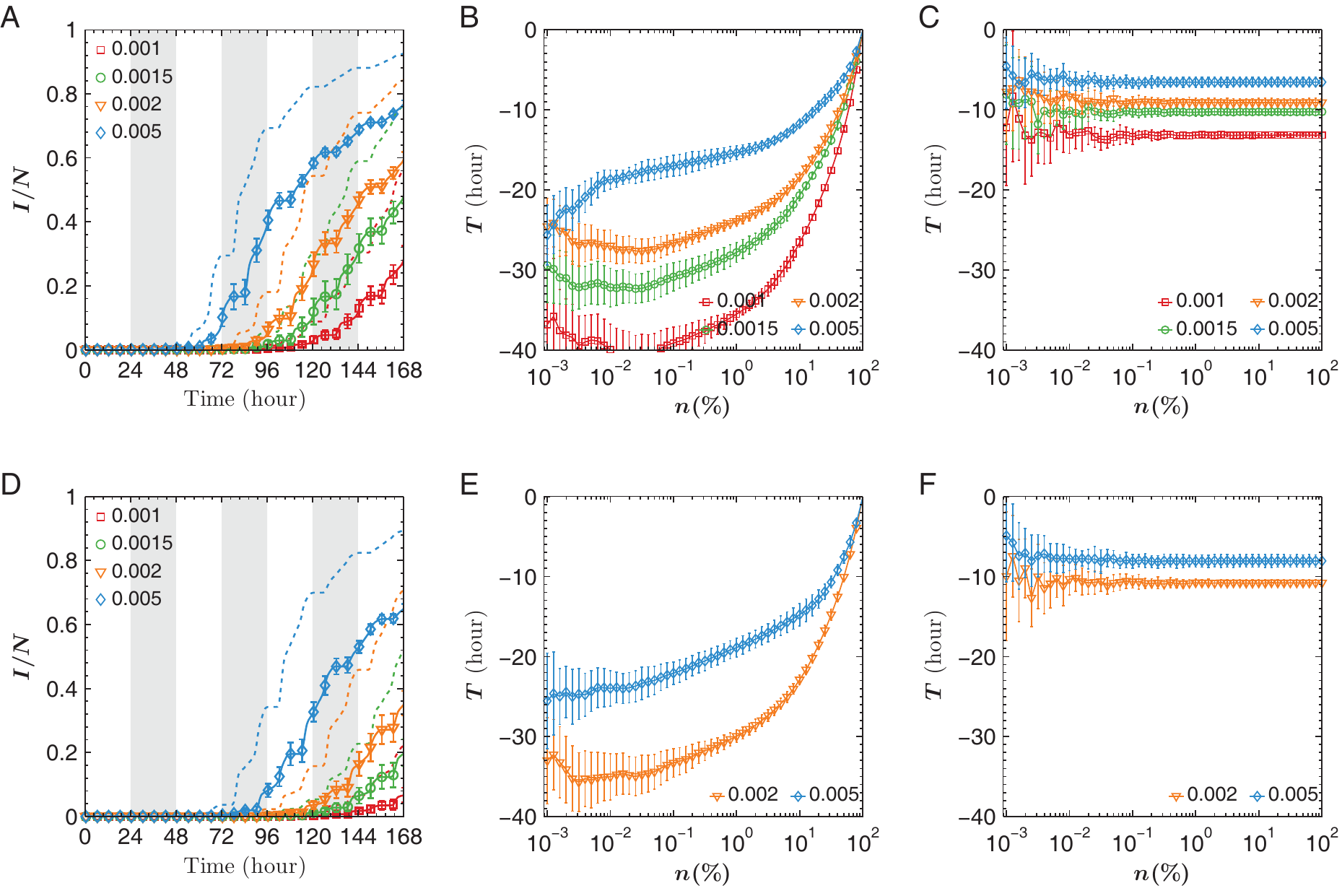}
\caption{Sensitivity on exposed duration (10 index cases). (A-C) Temporal ratio of infected cases (A), lead-time provided by sensors with highest degree (B) and friend sensors (C) when exposed duration is 6 hours. (D-F) Exposed duration is 12 hours. In this case, the spreading cannot reach 25\% when $\beta=0.001,0.0015$ after the whole period, so lead-time is not available. }
\label{figs6}
\end{figure*}

\end{document}